\PassOptionsToPackage{table}{xcolor}
\documentclass[sigconf]{acmart}
\usepackage{booktabs} 
\usepackage{multirow}
\usepackage{geometry}
\usepackage{array}
\usepackage{CJKutf8}
\usepackage{epigraph}
\usepackage{pdflscape}     
\usepackage{booktabs}      
\usepackage{tabularx}      
\usepackage{array}         
\usepackage{ragged2e}      
\usepackage{makecell}      
\usepackage{placeins}
\usepackage{graphicx}
\usepackage{subcaption}
\usepackage[normalem]{ulem} 

\newcolumntype{Y}{>{\RaggedRight\arraybackslash}X}
\definecolor{rowgray}{gray}{0.96}

\AtBeginDocument{%
  }

\setcopyright{acmlicensed}
\copyrightyear{2018}
\acmYear{2018}
\acmDOI{XXXXXXX.XXXXXXX}
\acmConference[Conference acronym 'XX]{Make sure to enter the correct
  conference title from your rights confirmation email}{June 03--05,
  2018}{Woodstock, NY}
\acmISBN{978-1-4503-XXXX-X/2018/06}

\copyrightyear{2026}
\acmYear{2026}
\setcopyright{cc}
\setcctype{by}
\acmConference[CHI '26]{Proceedings of the 2026 CHI Conference on Human Factors in Computing Systems}{April 13--17, 2026}{Barcelona, Spain}
\acmBooktitle{Proceedings of the 2026 CHI Conference on Human Factors in Computing Systems (CHI '26), April 13--17, 2026, Barcelona, Spain}
\acmPrice{}
\acmDOI{10.1145/3772318.3791139}
\acmISBN{979-8-4007-2278-3/2026/04}

\begin{document}

\title[Technology-Enabled Language Choice]{Inclusive Mobile Learning: How Technology-Enabled Language Choice Supports Multilingual Students}

\author{Phenyo Phemelo Moletsane}
\email{pmoletsa@andrew.cmu.edu}
\orcid{0009-0005-9999-1725}
\affiliation{%
  \institution{Human-Computer Interaction Institute, Carnegie Mellon University}
  \city{Pittsburgh}
  \state{Pennsylvania}
  \country{USA}
}
\author{Michael W. Asher}
\email{masher@andrew.cmu.edu}
\orcid{0000-0002-1006-8813}
\affiliation{%
  \institution{Human-Computer Interaction Institute, Carnegie Mellon University}
  \city{Pittsburgh}
  \state{Pennsylvania}
  \country{USA}
}

\author{Christine Kwon}
\email{ckwon2@andrew.cmu.edu}
\orcid{0009-0001-2825-2280}
\affiliation{%
  \institution{Human-Computer Interaction Institute, Carnegie Mellon University}
  \city{Pittsburgh}
  \state{Pennsylvania}
  \country{USA}
}

\author{Paulo F. Carvalho}
\email{pcarvalh@andrew.cmu.edu}
\orcid{0000-0002-0449-3733}
\affiliation{%
  \institution{Human-Computer Interaction Institute, Carnegie Mellon University}
  \city{Pittsburgh}
  \state{Pennsylvania}
  \country{USA}
}

\author{Amy Ogan}
\email{aeo@andrew.cmu.edu}
\orcid{0000-0003-2671-6149}
\affiliation{%
  \institution{Human-Computer Interaction Institute, Carnegie Mellon University}
  \city{Pittsburgh}
  \state{Pennsylvania}
  \country{USA}
}

\renewcommand{\shortauthors}{Moletsane et al.}
\begin{abstract}
Most learners worldwide are multilingual, yet implementing multilingual education remains challenging in practice. EdTech offers an opportunity to bridge this gap and expand access for linguistically diverse learners. We conducted a quasi-experiment in Uganda with 2,931 participants enrolled in a non-formal radio- and mobile-based engineering course, where learners self-selected instruction in Leb Lango (a local language), English, or a Hybrid option combining both languages. The Leb Lango version of the course was used disproportionately by learners from rural areas, those with less formal education, and those with lower prior knowledge, broadening participation among disadvantaged learners. Moreover, the availability of Leb Lango instruction was associated with higher active participation, even among learners who registered for English instruction. Although Leb Lango learners began with lower performance, they demonstrated faster learning gains and achieved comparable final examination outcomes to English and Hybrid learners. These results suggest that providing local language options to learners is an effective way to make EdTech more accessible.
\end{abstract}



\maketitle
\section{Introduction}

The world is multilingual. According to UNESCO, at least half of the global population is bilingual, speaking two or more languages or dialects \cite{unesco_languages_2025}. Although precise estimates are hard to establish \cite{byers-heinlein_case_2019}, research suggests that roughly 50\% of the world’s population uses more than one language daily \cite{grosjean_bilingual_2010, bialystok_bilingualism_2012}. Thus, multilingualism and educational practices that use two or more languages are common in everyday life around the world \cite{dutcher_use_1994}. 
Despite the prevalence of multilingualism, implementing multilingual education remains a challenging endeavor, with significant social, economic, and linguistic consequences. \cite{tupas_inequalities_2015, probyn_language_2006, brock-utne_language_2004, meier_multilingual_2017, skutnabb_todays_2020}. Chief among these challenges is the fact that traditional instruction cannot feasibly be delivered in more than one language simultaneously, and learners generally do not have the option to choose the language of instruction. Digital learning environments are uniquely positioned to address this gap: they can deliver the same learning content in multiple languages and allow learners to choose or even alternate between languages based on their comfort or linguistic needs.

In this paper, we investigate the opportunities and consequences afforded by a multilingual educational technology (EdTech) platform. We focus on a non-formal education setting in a highly multilingual context: rural Uganda. This setting is particularly appropriate for this study because English is the official language and the common language of instruction that digital platforms would typically default to, while most learners primarily use local languages in their daily lives. We specifically investigate how offering language choice beyond English in a multilingual, radio- and phone-based Science, Technology, Engineering, and Mathematics (STEM) course relates to course participation and access in a non-formal learning environment.

We conducted a quasi-experimental study in 2024, in which learners self-selected their preferred language for the entire course: Leb Lango (a local language), English,  or a combination of both Leb Lango and English (referred to as Hybrid). We employed a quantitative approach to examine the following research questions:

\begin{itemize}
\renewcommand\labelitemi{}
    \item \textbf{RQ1:} What differences in course accessibility exist for disadvantaged learners who selected English, Leb Lango, or Hybrid instruction?
    \item \textbf{RQ2:} Do learners who selected Leb Lango, English, or Hybrid instruction start the course with different levels of prior knowledge?
    \item \textbf{RQ3:} How do learners who selected Leb Lango, English, or Hybrid instruction engage with the lesson content?
    \item \textbf{RQ4:} How do learners who selected Leb Lango, English, or Hybrid instruction progress through the course?
    \item \textbf{RQ5:} How do learners who selected Leb Lango, English, or Hybrid instruction perform in the course?
\end{itemize}

Our findings indicate that providing a digital technology environment that allows language choice, broadened participation among learners with the least formal education and the lowest baseline knowledge. Although learners who chose Leb Lango instruction entered with lower prior knowledge, they demonstrated higher engagement and faster learning gains than English learners throughout the course, and achieved comparable final examination outcomes to those who selected English and Hybrid instructions. Our contributions are threefold: (1) we conceptualize language choice as a context-sensitive way to personalize digital learning and expand education access to learners who might otherwise be excluded from English-only instruction; (2) we provide the first large-scale quasi-experimental evidence on multilingual EdTech in resource-constrained rural settings, highlighting the nuanced ways in which language choice could benefit different learners; and (3) we outline design implications for EdTech designers and instructors on the considerations of implementing language support to better serve learners in multilingual contexts.

\section{Background and Related Work}

As the need to support education in low-infrastructure communities gains increasing attention, research on mobile learning has grown substantially. Researchers are exploring how mobile devices can create new learning opportunities in contexts where traditional educational resources are limited. To situate our study, we review several bodies of prior work. First, we cover research on mobile learning in low-resource, non-formal settings, including radio- and phone-based interventions. We then turn to alternative models of distance and remote education, particularly Massive Open Online Courses (MOOCs), which inform our understanding of access through the lens of linguistic inclusion. Although MOOCs differ from our low-tech context, this body of work highlights why linguistic inclusion matters and helps frame the importance of incorporating multilingualism in mobile learning. Finally, we examine research on multilingualism in mobile learning and broader digital learning environments. Together, these strands of research position our study and help highlight the gaps it addresses. In this paper, we use "low-resource" or "resource-constrained" to refer to settings with significant limitations in financial, human, and infrastructural resources, and "non-formal learning" to refer to any learning that occurs outside the formal school system.

\subsection{Mobile Learning in Rural and Low-Resource Contexts}

There exists a body of related work pertaining to taking advantage of mobile learning technologies to improve educational access and outcomes in resource-constrained environments. Studies indicate that mobile phones, including both smartphones and basic cellular devices, are the primary tools used by students and instructors for mobile learning \cite{kaliisa_systematic_2017}. A substantial body of research shows that integrating mobile technologies into classroom instruction can broaden learning opportunities in low-resource settings, for example, by enabling remote co-teaching, strengthening instructional support, and improving learners’ knowledge acquisition \cite{frias-martinez_mobilizing_2012, guo_remote_2022, wahyuni_effectiveness_2024, nestel_evaluation_2010, breazeal_mobile_2016}. Large-scale deployments further demonstrate the potential of mobile technologies: Breazeal et al. (2016) provided low-cost tablets with early literacy content for young children in under-resourced communities across Ethiopia, Uganda, South Africa, India, and rural United States, where children showed measurable pre- to post-intervention gains in literacy, and those with tablet access generally outperformed peers who did not receive tablets \cite{breazeal_mobile_2016}. 

However, much of this work remains situated within formal classroom settings. Many learners in rural areas and resource-constrained environments face particularly severe barriers to schooling: long and unsafe travel to school, a shortage of trained teachers, scarce instructional materials, poor access to technology, inadequate infrastructure, and economic demands that force children into household labor, often interrupting or preventing school attendance \cite{irvin_educational_2012}. For example, research highlights rural learners' significant marginalization in STEM education, which limits their development of STEM skills and pursuit of STEM careers \cite{mutambara_determinants_2021, sung_enhancing_2025, harris_stem_2018}. Given these structural challenges, non-formal learning supported by mobile technologies has become an important complement to formal schooling in rural and low-resource environments \cite{irvin_educational_2012, kukulska-hulme_designing_2007}.

A related stream of research has examined how mobile technologies can support learning outside traditional classroom settings \cite{kumar_exploratory_2010, kizilcec_mobile_2021, hawi_leveraging_2021, kwon_investigating_2024, intarat_icts_2017}. In evaluating the impact of mobile technologies in low-resourced and out-of-school contexts, prior research has used outcomes such as motivation, engagement, and performance, with some studies reporting improvements relative to learners’ baseline performance and, in certain cases, compared to peers without access to the technology \cite{breazeal_mobile_2016, kwon_investigating_2024, khan_mitigating_2019}. For example, Kizilcec et al. (2021) examined the impact of a text message-based system delivering lessons and quizzes to learners in Kenya, and a voice-based platform providing recorded lessons, quizzes, and feedback to support literacy development in Côte d’Ivoire. Their study found that learners continued to engage with these phone-based technologies to supplement their formal education during periods of disruption, largely due to the affordability of text and voice messaging \cite{kizilcec_mobile_2021}. Similarly, Kumar (2010) conducted a 26-week study in rural India that pre-loaded mobile phones with games targeting English (as a Second Language) vocabulary skills to investigate how children use cellphones when unsupervised in their everyday lives; and found that children were motivated to use the phones and achieved measurable learning outcomes in vocabulary, though infrastructural constraints such as electricity and gender norms limited access \cite{kumar_exploratory_2010}. Further, some studies indicate that in rural and low-resource settings, simpler mobile tools such as basic phones and radios are particularly beneficial, as they are more widely available, affordable, and better suited to local infrastructure than high-bandwidth technologies that require stable connectivity and advanced hardware \cite{kwon_investigating_2024, valk_using_2010}. 

Radio, in particular, has long served as an effective, low-cost educational technology with well-documented benefits in low-infrastructure settings \cite{bosch_interactive_1997}. Interactive radio instruction, which combines radio broadcasts with guided learning prompts, has been shown to improve instructional quality and learning outcomes in contexts such as Nigeria, and has even been adapted to deliver distance learning to rural and nomadic communities without access to formal schooling \cite{aderinoye_integrating_2007}. More recent work in rural Uganda shows that combining radio with mobile phones can support large-scale STEM learning and increase motivation among learners who typically face restricted educational opportunities \cite{kwon_investigating_2024}. Building on this evidence, our work adopts a similar radio-and-phone-based model designed for low-resource environments. Prior studies investigating radio and phone interventions have given limited attention to how such systems accommodate learners’ linguistic contexts, which represents a critical gap in rural Uganda, where language diversity is high. Learning in these communities is experiential and socially situated, with local languages playing a central role in knowledge sharing. Moreover, learners in rural areas often have fewer opportunities to practice English than their urban peers, contributing to persistent literacy and comprehension gaps \cite{hossain_english_2016}. Despite language being a central mediator of access, participation, and comprehension, multilingualism remain underexplored in EdTech research \cite{uchidiuno_designing_2018, ruiperez-valiente_large_2022, grandon_factors_2005, liu_mooc_2016, traxler_introduction_2014}.

\subsection{Linguistic Recognition in MOOCs}

The need for multilingual learning has been well documented in MOOCs \cite{uchidiuno_designing_2018, ruiperez-valiente_large_2022, grandon_factors_2005, liu_mooc_2016, traxler_introduction_2014}, reflecting long-standing insights from classroom-based research. A substantial body of evidence from formal education highlights the value of incorporating local languages into curricula, with consistently stronger outcomes for learners taught in their mother tongue across diverse contexts, including the Philippines, Kenya, South Africa, and the United States \cite{walter_mother_2011, piper_implementing_2016, trujillo_use_2020, thomas_national_2002, nomlomo_science_2007}. For example, a medium-scale randomized controlled trial (RCT) in Kenya showed that incorporating mother tongue instruction into early math and reading programs significantly improved literacy outcomes, particularly oral fluency and comprehension, with effect sizes on mother tongue literacy averaging between 0.3 and 0.6 standard deviations \cite{piper_implementing_2016}.
In the context of digital learning, including MOOCs, much of the research broadly highlights the importance of language, culture, and social context in shaping learners’ access to and engagement with the learning material. Warschauer (2003) argues that meaningful access to Information and Communication Technology (ICT) requires far more than providing devices or connectivity; content, language, literacy, and community structures must all be considered \cite{warschauer_technology_2003}. UNESCO also notes that mobile learning content often lacks relevance due to limited language options and a scarcity of culturally grounded material \cite{west_policy_2013}. Empirical studies highlights several language-related barriers in digital learning contexts. Studies demonstrate that non-native English speakers engage differently with educational videos \cite{uchidiuno_designing_2018}; that learners in multilingual contexts such as Bangladesh, Nepal, Senegal, and Sudan routinely navigate multiple languages while facing constraints in English proficiency that limit technology use \cite{kukulska-hulme_roles_2023}; and that limited literacy in both English and digital technologies remains a major barrier to MOOC participation in developing countries \cite{liyanagunawardena_impact_2013}. Large-scale surveys of MOOC participants further show that learners find it easier to learn in their native language \cite{ruiperez-valiente_large_2022}. 

Despite broad recognition of the importance of language in EdTech, there is limited evidence on interventions that integrate multilingual support and examine their impact. Our work begins to address this gap by investigating the opportunities and consequences afforded by a multilingual digital educational technology platform. Nonetheless, a small but growing body of studies has begun to explore comparable approaches.

\subsection{Language Support and Multilingualism in EdTech}

Very few EdTech systems integrate multilingual support \cite{jantjies_multilingual_2012, jantjies_mobile_2013, jantjies_mobile_2015, castillo_early-grade_2019}. For example, Jantjies and Joy (2012, 2013) introduced bilingual mobile learning tools in South Africa that enabled high school students to access science and mathematics content (simplified notes, textbook notes) in both English and local languages; learners reported that bilingual access aided conceptual understanding. Similarly, a multilingual reading program in rural South Africa offering support in three local languages plus English produced substantial gains in reading fluency and comprehension \cite{castillo_early-grade_2019}. Mohammed and Mohan (2013; 2015) studied cultural localization in Trinidad and Tobago, developing a localized intelligent tutoring system (CRIPSY/CRITS) that taught programming using English-based Creoles, and found that students preferred localized content and even requested to adjust the density of cultural and linguistic context \cite{mohammed_case_2013, mohammed_dynamic_2015}. Additional evidence on the role of language in digital learning comes from a multilingual MOOC study examining an online course offered in both English and Arabic, which found that participants across both versions displayed similar intrinsic motivations and self-determination, underscoring that multilingual delivery can support diverse learner needs \cite{barak_motivation_2016}. However, these studies examine classroom-based interventions and lack flexible language options - a key distinction of our study. Flexible language choice is particularly important for supporting learners with varying proficiencies and comfort levels, especially in non-formal learning environments.
Outside formal schooling, efforts to incorporate linguistic inclusion have largely focused on language learning itself (e.g., mobile-assisted language learning) \cite{paredes_loch_2005, ahmad_impact_2017, ahmad_integrating_2019}, rather than on delivering disciplinary content in local or multiple languages.

Overall, across all these bodies of work, several important gaps remain. Prior radio- and phone-based learning interventions have not examined how such systems should accommodate learners’ linguistic contexts, despite their widespread use in low-resource environments. Existing multilingual EdTech research has focused primarily on formal classroom settings, limiting its relevance for non-formal learning contexts where local languages dominate everyday communication and language proficiency variations are most pronounced. Most mobile learning initiatives outside school settings have typically incorporated language accommodations only for language learning, rather than for the delivery of disciplinary content; our work extends this scope by offering STEM instruction in a local language. Moreover, even in studies that have implemented multilingual EdTech, learners have not been given the option to choose their instructional language. There is no large-scale quantitative evidence from sub-Saharan Africa on how language preference supports participation, engagement, or learning outcomes in low-tech mobile learning environments. Our study addresses these gaps by examining how multilingual delivery of a STEM course in a large-scale, non-formal mobile learning system supports access and participation.

\section{Methods}

We adopted a quasi-experimental design for this study in collaboration with the educational technology company, Yiya Solutions, which develops and delivers practical, remote engineering education courses to rural learners in Northern Uganda. We analyzed data from a course delivered between August 5 and November 24, 2024. The recruitment, consent and data collection of participants were conducted entirely by our collaborators in Uganda, and the authors of this paper were not involved in these processes. The study received approval from a relevant Ugandan ethics committee.
\subsection{Study Context and Recruitment}

The study was conducted in Lira District, a predominantly rural district in Northern Uganda. Uganda is home to over 60 indigenous languages \cite{criper_linguistic_1971}, in addition to two official but non-indigenous languages, English and Kiswahili \cite{schmied_varieties_2008}. The country’s Language-in-Education policy (LiEP) aims to support bilingual learning, particularly by providing local language instruction in government schools  \cite{altinyelken_dilemmas_2014, ssentanda_challenges_2014}, where local languages serve as the medium of instruction during the first three years of primary school, after which English becomes the primary language of instruction from Primary Four through secondary and higher education \cite{mulumba_challenges_2012}. The population in the Lira District is largely from the Lango ethnic group, and the primary language spoken is Leb Lango, a minority language used by approximately 5\% of the Ugandan population \cite{ubos_lango_profiles_2025}. Many people in the district come from low socio-economic backgrounds, where limited financial resources restrict access to education. 

The learners were recruited for the remote course through radio advertisements that provided instructions on how to access the program. They were asked to complete a recruitment questionnaire via text message and given the option to consent to enroll and allow their learning data to be used for course improvement and research purposes. Out-of-school learners were actively encouraged to participate in the program. 

All learners indicated their language preference during registration and were assigned to a group based on this choice. They were offered three options for instruction: English, one of the official languages in Uganda; Leb Lango, a local language spoken in the district; and a Hybrid option, which provided the flexibility to choose or alternate between English and Leb Lango at the start of each course unit. 

\subsection{Participant Demographics}
Table~\ref{tab:demographics} presents the demographic characteristics of the learners who participated in the study. We recruited a total of 2,931 learners for the study, with a fairly balanced gender distribution of men and women. In terms of educational attainment, the largest group of learners had completed upper primary education (P5-P7) or lower secondary education (S1-S4). Only a small proportion of the learners in the sample had completed upper secondary education (S5 or above) or very little schooling (lower primary: P1-P4). The sample included learners who were currently enrolled in school and those who were out of school. The age distribution was broad, with many learners in their late teens and early twenties, and fewer younger adolescents. Most participants did not have children. The majority reported living in villages, with others residing in cities, town centers, and a very small fraction in refugee settlements. Overall, this demographic profile reflects a predominantly rural sample, a balance of gender, a mix of in- and out-of-school learners, and a wide age range, illustrating the diversity of learners targeted by the EdTech program.
\begin{table*}[ht]
\centering
\caption{Demographic characteristics of study participants. Percentages are calculated within each category.}
\Description{This table shows the demographic characteristics of study participants. Percentages are calculated within each category.}
\label{tab:demographics}
\begin{tabular}{l l c}
\hline
\textbf{Characteristic} & \textbf{Category} & \textbf{Percentage (\%)} \\
\hline
\textbf{Gender} & Man & 54 \\
                & Woman & 46 \\
\hline
\textbf{Highest Education Level} & P1-P4 (Lower primary) & 9 \\
                                 & P5-P7 (Upper primary) & 37 \\
                                 & S1-S4 (Lower secondary) & 44 \\
                                 & S5 or above (Upper secondary) & 10 \\
\hline
\textbf{Current School Enrollment} & In-school & 45 \\
                                   & Out-of-school & 55 \\
\hline
\textbf{Age Group} & 11 years and below & 7 \\
                    & 12-15 years & 11 \\
                    & 16-19 years & 28 \\
                    & 20-24 years & 36 \\
                    & 25+ years & 18 \\
\hline
\textbf{Parental Status} & No children & 73 \\
                         & Has children & 27 \\
\hline
\textbf{Neighborhood} & Village & 60 \\
                      & Town center & 12 \\
                      & City & 27 \\
                      & Refugee settlement & <1 \\
\hline
\end{tabular}
\end{table*}

\subsection{The Educational Platform: Yiya Solutions and Learning Content}

\begin{figure}[h]
  \centering
  \includegraphics[width=\linewidth]{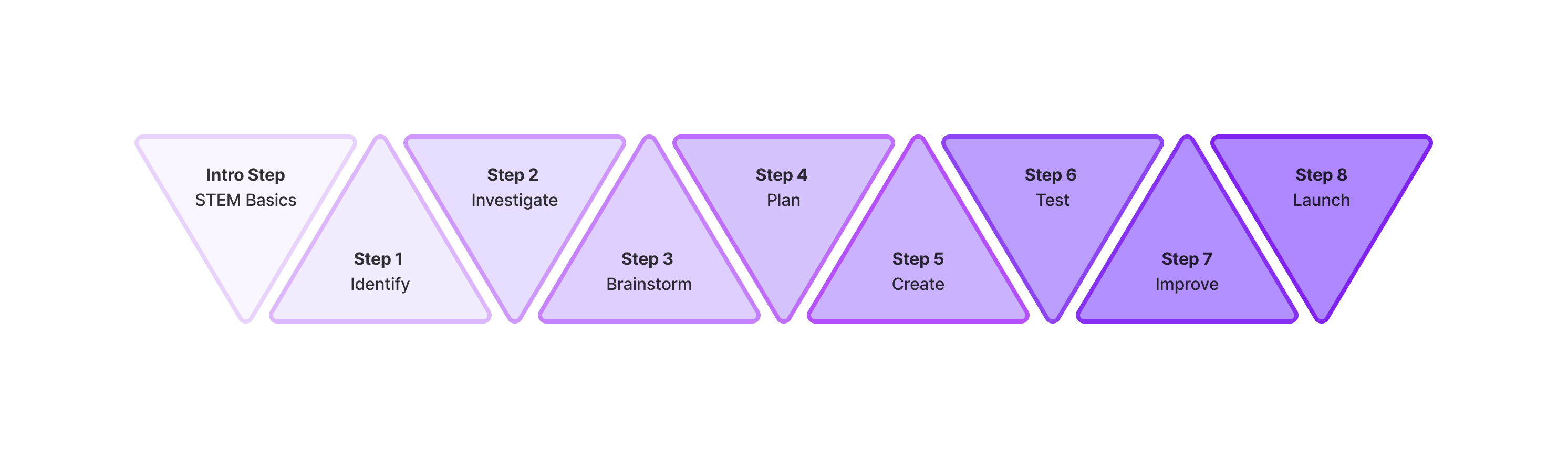}
  \caption{Engineering Process Design steps in the Yiya Solutions course curriculum. The curriculum consists of nine ordered steps to create the final product. Each step is made of several lessons: Introduction Step: Basics of STEM education (4 lessons); Step 1: Identify (3 lessons); Step 2: Investigate (4 lessons); Step 3: Brainstorm (4 lessons); Step 4: Plan (4 lessons); Step 5: Create (4 lessons); Step 6: Test (3 lessons); Step 7: Improve (6 lessons); Step 8: Launch (4 lessons).}
  \Description{This image depicts an overview of the steps in the course curriculum. The steps follows the engineering design process to create the final. product. The curriculum consists of nine ordered steps to create the final product. Each step is made of several lessons: Introduction Step: Basics of STEM education (4 lessons); Step 1: Identify (3 lessons); Step 2: Investigate (4 lessons); Step 3: Brainstorm (4 lessons); Step 4: Plan (4 lessons); Step 5: Create (4 lessons); Step 6: Test (3 lessons); Step 7: Improve (6 lessons); Step 8: Launch (4 lessons).} 
  \label{fig:curriculum}
\end{figure}

Yiya Solutions is an offline, remote learning platform designed to bring interactive hands-on 15-week STEM education to participants in rural Uganda, where access to the Internet, smartphones, academic materials and computers is often limited \cite{yiya_home_2025}. As such, the learning platform is low-tech and relies on widely available tools in rural Africa to reach remote learners: basic keypad phones and radios.
The course curriculum followed the steps of the Engineering Design Process (EDP) and was structured into nine ordered steps (or units): Introduction Step: Basics of STEM education; Step 1: Identify; Step 2: Investigate; Step 3: Brainstorm; Step 4: Plan; Step 5: Create; Step 6: Test; Step 7: Improve; Step 8: Launch. Each of the step is made of several lessons as shown in Figure~\ref{fig:curriculum}. The EDP steps are integrated into the weekly lessons to give learners hands-on, problem-solving experience. Learners are guided to build practical technologies that address everyday challenges in their communities through scientific experiments using locally available materials. In the version of the course we studied, the learners were taught to create a solar-powered food dehydrator, a device that uses solar energy to dry and preserve produce, such as fruits and vegetables. This project is particularly relevant, as communities in this region primarily rely on subsistence farming, making the dehydrator a practical solution to improve food preservation. The completed product serves as the learners’ final output, demonstrating both their STEM knowledge and practical skills.

\subsection{Instructional Language}
The curriculum content was identical across both languages. Multiple translators independently produced Leb Lango translations, which were subsequently validated through comparison and consensus-building to ensure consistency across the curriculum. Translators noted that some English STEM terms lacked widely used local equivalents, leading them to adopt more complex or less common Leb Lango terms. These choices reflected an intentional trade-off: maintaining scientific correctness without oversimplifying concepts. 

All assessments, lesson content, and radio instructions were delivered in the learners’ chosen language option at registration. Radio broadcasts were scheduled on the same day but at different times: English sessions at 12:00 PM EAT and Leb Lango sessions at 2:00 PM EAT. Learners in the Hybrid option could switch languages on their phones through a scheduled language switch feature at the end of each step or within a step by updating the language in the profile menu.

\subsection{Course Delivery, Activities and Engagement}

The course consists of lessons, practice questions, weekly assessments, a final examination, and a hands-on project. All questions in the course are primarily in a multiple-choice format and are delivered in the learners’ chosen language option. Once registered, learners receive alerts via Robocalls and Short Message Service (SMS), which provide information about upcoming content and any materials they will need. Learners are expected to prepare the materials and related assignments prior to the lesson. The instruction and content of the lesson are delivered through radio broadcasts by an instructor.
To engage with the lesson content, on specific days, learners tune in to a dedicated local radio broadcast accessed from home or anywhere in their community, where the instructor guides them through the educational content and the EDP steps to build the final product. Each broadcast lasts approximately 30 minutes, and each consecutive week focuses on a specific step of the course curriculum. During lessons, instructors engage learners by asking practice questions via the radio, which learners answer in real time using Unstructured Supplementary Service Data (USSD) menus on their phones or via live call-ins. USSD is a communication protocol that mobile network operators use to send text messages between a mobile phone and an application server for free. It allows learners to submit assessment responses without typing, making it ideal for multiple-choice questions and constrained settings. Practice questions can also be accessed outside of the radio broadcasts. The course emphasizes hands-on projects and problem-solving rather than formulas or expressions, enabling learners to apply scientific concepts in contextually meaningful ways. For example, a sample question from Step 4 reads:  "\textit{All these materials have properties that let sunlight pass through and can hold air inside EXCEPT: 1. Cling film; 2. Glass; 3. Black polyethylene bag; 4. Hard clear plastic".} Learner responses from the USSD are logged, allowing tracking  of learner engagement, progression, and performance. To encourage sustained engagement, the platform employs several strategies. For example, each radio lesson includes a prize code that the instructor reads aloud;  learners must enter the code on their phones before accessing the lesson’s practice questions. Learners are also incentivized for attentive listening and accurate responses; for example, those who actively participate may receive materials kits to support their final products.

\begin{figure}[h]
  \centering
  \includegraphics[width=\linewidth]{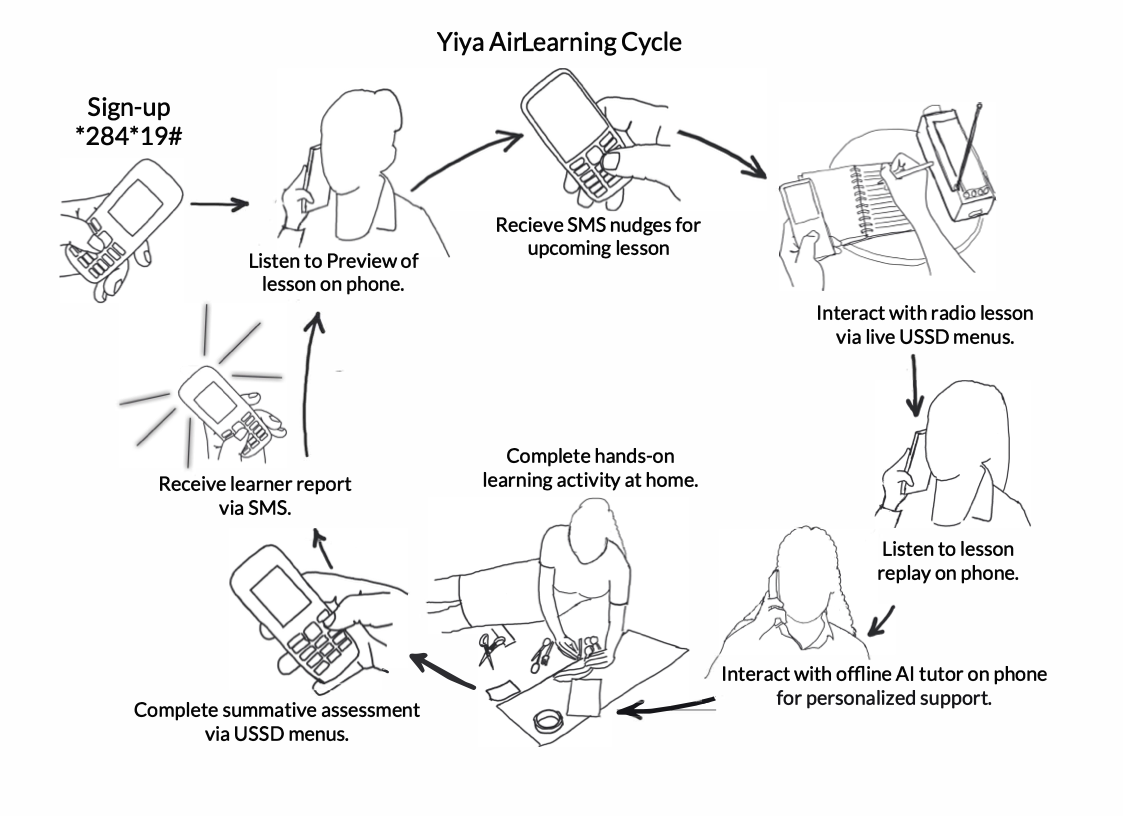}
  \caption{A visual presentation of how learners interact with the Yiya Solutions remote course. Learners access the course and review the course material by dialing the USSD code on their phones. On certain days of the week, learners will tune in to live radio broadcast delivered in the language they selected. Instructors deliver the lesson content, activities, and questions through radio. Learners then interact with the lesson content and questions via USSD on their phones.}
  \Description{This image depicts a cycle illustrating how learners interact with radio broadcasts and mobile phones to access the course. The cycle highlights key elements including signing up via USSD, listening to lessons via radio, reading SMS directions, taking quizzes on mobile phones and carrying out hands-on experiments at home.}
  \label{fig:student_interaction}
\end{figure}

Learners also complete weekly multiple-choice assessments via USSD, aligned with the content from the previous week’s lessons. At the end of the program, they take a final 16-question multiple-choice assessment, delivered through radio, and answered via USSD. In addition to assessments, learners design and build an innovative technological product related to the course topic (solar-powered food dehydrator). Participants who achieve a score of 75\% or higher receive a certificate that recognizes both their theoretical knowledge and their practical skills. Figure~\ref{fig:student_interaction} summarizes the visual cycle of participants' interaction with the course. The course concludes with a community exhibition, where the learners showcase their final products, demonstrating the real-world applicability of their learning.


\subsection{Potential Challenges and Considerations}

While the learning platform reaches learners in remote rural communities, several challenges and constraints may emerge in its delivery. Frequent electricity cuts which are common in these regions can disrupt phone use, and some learners may not own personal mobile phones, resulting in device sharing that can reduce engagement. In such cases, Robocall and SMS notifications may not always be received which can delay learners’ preparation for lessons. Radio-based instruction, while widely accessible, is linear and broadcast-dependent, limiting opportunities for real-time interaction for all learners simultaneously. Finally, USSD-based responses restrict interaction to multiple-choice formats due to character limits, making open-ended or more expressive responses impractical in any language. However, this constraint aligns with the course’s design, which emphasizes hands-on projects and problem-solving rather than written explanations.

\subsection{Data and Measures}
Our study included several measures to capture learner characteristics, behavior, and outcomes. We captured learners' demographic information (highest education level, age group, neighborhood type, gender, current school enrollment, and parental status). We also recorded the learners' language choice: Leb Lango, English, or a Hybrid option as their instructional language. 

Prior knowledge was measured using the learners’ test scores from the first week of the course. This was viewed as a particularly context-relevant approach to measurement: rural, out-of-school learners who see a test as their first experience with a platform are very likely to drop out, which could result in high attrition among those with low prior knowledge.

Performance was assessed using the proportion of correct responses on multiple-choice practice questions, weekly tests, and the final examination. We measured engagement through prize code accuracy, which indicates whether the learners correctly submitted the code read aloud during each radio broadcast (see \cite{asher_validating_2025} for validation of this method), and by identifying which broadcast each learner listened to, based on the distinct prize codes used for the English and Leb Lango radio broadcasts.

\section{Analyses and Results}
In this section, we outline the analyses and report the results for each research question. We conducted all analyses in R (Version 4.5.1, R Core Team, 2025).

\subsection{RQ1: What differences in course accessibility exist for disadvantaged learners who selected English, Leb Lango, or Hybrid instruction?}
 
We first investigated how the demographic characteristics of the learner relate to the language preference given three options: Leb Lango, English, and Hybrid. Understanding this relationship is important for identifying which groups benefit the most from local language instruction and how remote courses can better support learners in these contexts. We fitted a multinomial logistic regression model with language choice as the dependent variable. We used Leb Lango as the baseline category to which the other language options are compared. Learners who registered and chose a language option but had missing values from not attempting any course steps were excluded from this analysis. The final sample for this analysis included 2,924 learners who: (1) registered for the course, (2) selected a language preference at registration, and (3) attempted at least one step in the course. All our independent variables were categorical and included: Highest education level, Neighborhood type (collapsed into "Town or city" versus "Village or settlement"), Current school enrollment, Gender, and Parental status. Because age was moderately correlated with education (Cram\'er\textquotesingle s $V > 0.30$
), we excluded it from the model. We assessed model fit for the multinomial logistic regression using McFadden’s pseudo $R^2$ and a likelihood ratio test. The model was statistically significant, $\chi^2 = 571.26, \ p < .001$, indicating that the demographic variables improved fit. McFadden’s $R^2 = 0.091$ suggests a modest explanatory power, with the predictors accounting for roughly 9\% of the variation in language preference.

\begin{figure}[h]
    \centering
    \begin{center}
    \begin{subfigure}[b]{1\linewidth}
        \centering
        \includegraphics[width=\linewidth]{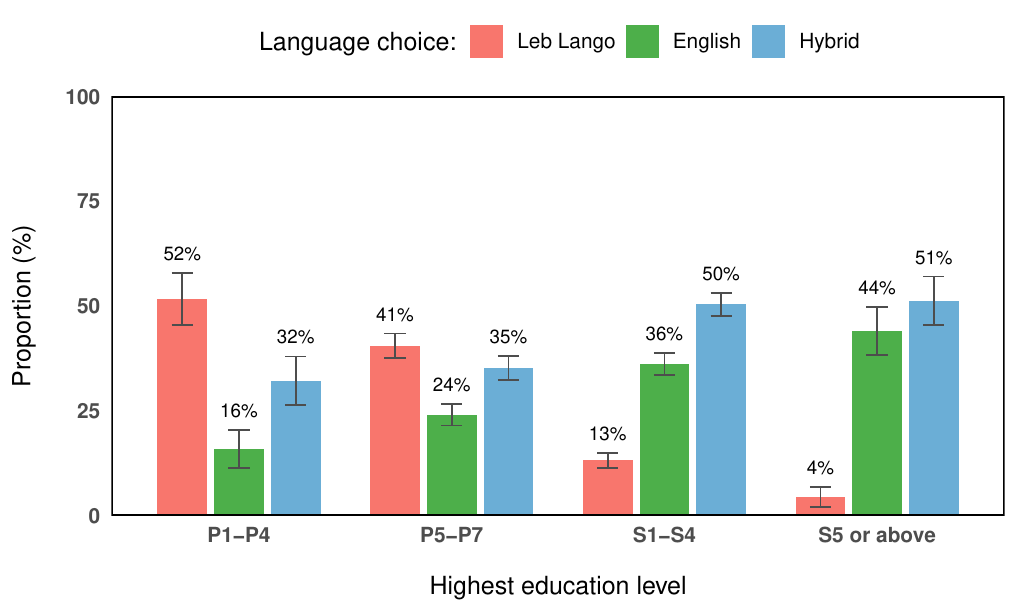}
        \caption{}
        \label{fig:language_ed}
    \end{subfigure}
    \par\vspace{0.5cm}
    \begin{subfigure}[b]{1\linewidth}
        \centering
        \includegraphics[width=\linewidth]{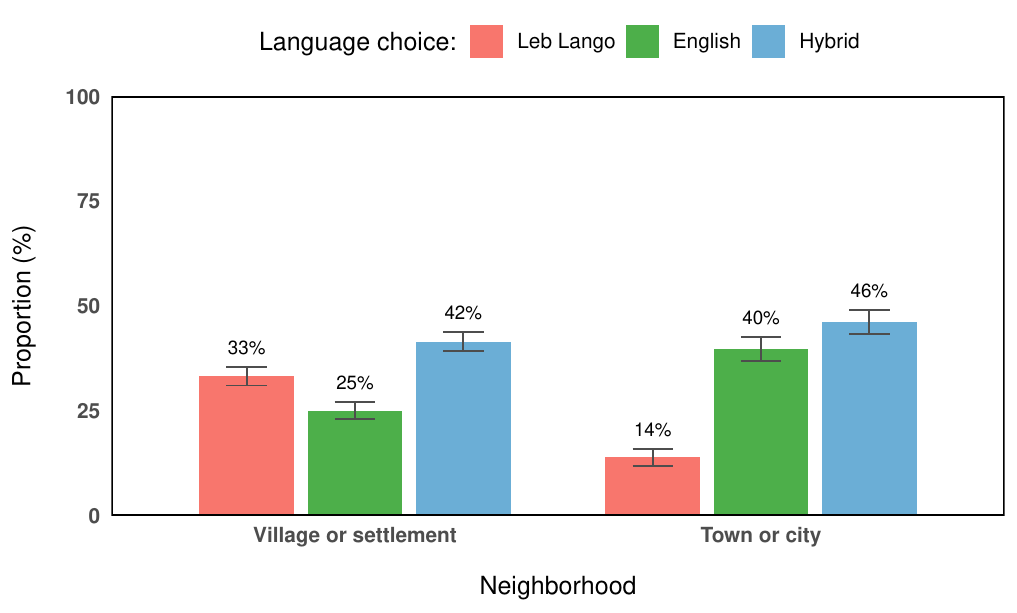}
        \caption{}
        \label{fig:language_ne}
    \end{subfigure}
    \caption{Proportion of learners selecting each language option by demographic characteristics. Figure (a) shows language preference by highest education level, indicating how learners with lower and higher education levels chose between the available language options. Figure (b) shows language preference by neighborhood type, illustrating how learners from villages, town centers, cities, and refugee settlements distributed their choices among the language options.}
    \Description{This figure shows the proportion of learners selecting each language option across different demographic characteristics. Figure (a) shows language preference by highest education level, indicating how learners with lower and higher education levels chose between the available language options. Figure (b) shows language preference by neighborhood type, illustrating how learners from villages, town centers, cities, and refugee settlements distributed their choices among the language options.}
    \label{fig:language_demographics}
    \end{center}
\end{figure}

Upon registration for the remote course, 43\% (n = 1,269) of students selected the Hybrid condition, 31\% (n = 902) selected English, and 26\% (n = 753) selected Leb Lango. The local language option (Leb Lango) was chosen by the fewest learners. However, looking more closely at who these learners are, a more granular analysis reveals how linguistic choices interact with structural barriers faced by different groups of learners in this resourced-constrained setting. Figure~\ref{fig:language_ed} reveals that learners' language preferences differed by highest education attainment. Among learners in lower primary (P1-P4), 52\% chose Leb Lango. This proportion steadily decreased with higher education levels: 41\% in upper primary (P5-P7), 13\% in lower secondary (S1–S4), and 4\% in upper secondary (S5 or above). In contrast, the average proportion of learners who chose English and Hybrid increased with higher education attainment. The results of our multinomial logistic regression model further showed that the highest level of education was significantly associated with language preference. Compared to learners in lower primary (P1-P4), those in upper primary (P5-P7) were 2.60 times more likely to choose English over Leb Lango ($95\% \text{ CI } [1.73, 3.89],\ p < .0001$) and 1.63 times more likely to choose Hybrid over Leb Lango ($95\% \text{ CI } [1.18, 2.26],\ p < .0001$). Learners in lower secondary (S1-S4) had 11.29 times the odds of selecting English ($95\% \text{ CI } [7.44, 17.12],\ p < .0001$) and 6.83 times the odds of selecting Hybrid ($95\% \text{ CI }[4.85, 9.61],\ p < .0001$) over Leb Lango. At the highest level, learners with upper secondary (S5 or above) had 38.50 times the odds of selecting English ($95\% \text{ CI } [19.32, 76.73],\ p < .001$) and 19.25 times the odds of selecting Hybrid ($95\% \text{ CI } [10.12, 36.62],\ p < .0001$) compared to those with lower primary education level (P1-P4).

Similarly, language preferences varied by neighborhood (see Figure  \ref{fig:language_ne}). Learners living in villages or settlements were more likely to select Leb Lango (33\%), while 25\% chose English and 42\% opted for Hybrid. In contrast, learners who reside in urban areas were less likely to choose Leb Lango (14\%), with 40\% choosing English and 46\% choosing Hybrid. Neighborhood variable was significantly associated with language choice: compared to learners living in villages or settlements, those in urban areas were 3.09 times more likely to select English ($95\% \text{ CI } [1.85, 2.87],\ p < .0001$) and 2.31 times more likely to choose Hybrid ($95\% \text{ CI } [1.85, 2.87],\ p < .0001$) over Leb Lango. These patterns suggest that learners in urban areas are more inclined to select English or Hybrid instruction, whereas learners in rural villages show a stronger preference for Leb Lango.

In general, our results suggest a clear association between learner demographic characteristics and course language choice. Preference for Leb Lango is the highest among learners with the least schooling education levels and those living in villages or settlements. Although we focus on education level and neighborhood type here, other demographic factors, such as parental status and current school enrollment, exhibited patterns consistent with these trends.
\subsection{RQ2: Do learners who selected Leb Lango, English, or Hybrid instruction start the course with different levels of prior knowledge?}
\begin{figure}[H] 
\centering \includegraphics[width=\linewidth]{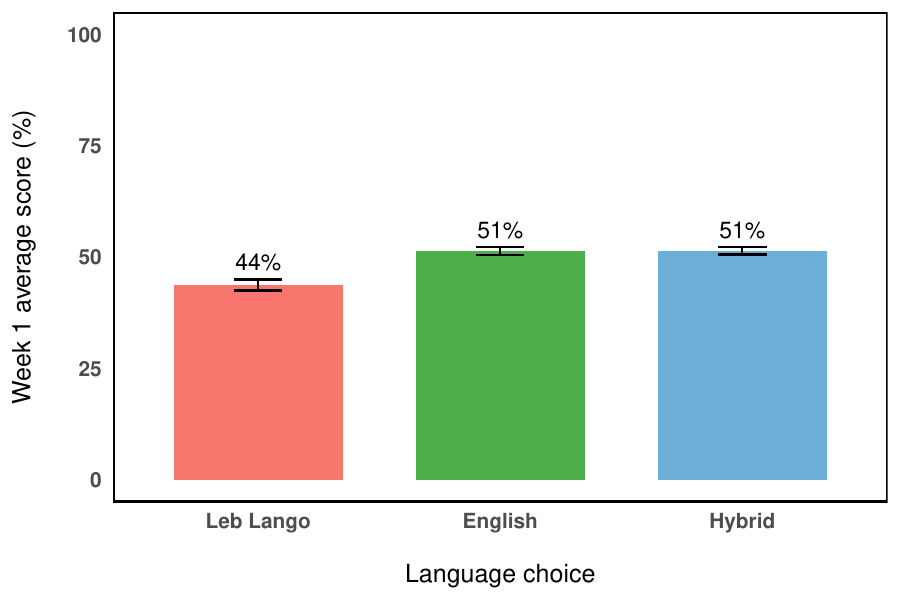} 
\caption{Average Week 1 score by language choice. Week 1 score is the learners’ average test scores from Week 1 as a proxy for their baseline performance.}
\Description{This figure shows the average Week 1 score for learners based on their language choice. It illustrates differences in baseline performance between learners who selected different language options for the course.}
\label{fig:week_lang} 
\end{figure}

For our second research question, we examined how learners' prior knowledge might relate to their choice of instructional language, hypothesizing that learners with stronger baseline performance may be more likely to select English or Hybrid instruction. This analysis was particularly important to understand whether lower-prior-knowledge learners systematically choose the local language and for exploring how language options may serve as a mechanism to improve access for more vulnerable learners. To address missingness, we excluded learners who did not attempt at least eight steps in the course, ensuring that our sample reflected those who participated in the full course. After this filtering, the analysis included 1,326 learners: 442 in English, 583 in Hybrid, and 301 in Leb Lango. Since no formal baseline scores were available, we used learners’ average test scores from Week 1 as a proxy for their early performance. For each learner, we calculated this average score by summing all test scores for Week 1 and dividing by the number of attempts, producing a single indicator of Week 1 performance. We then used a multinomial logistic regression model, using Leb Lango as the baseline category to which the other language options are compared. The model estimated the log-odds of selecting English or Hybrid compared to Leb Lango, with Week 1 performance as the independent variable. The results are shown in Figure~\ref{fig:week_lang}. 

Average week 1 scores were higher among learners who chose English (51\%) and Hybrid (51\%) compared to those who chose Leb Lango (44\%) (Figure~\ref{fig:week_lang}). Week 1 performance was significantly associated with language choice: learners with higher Week 1 scores were more likely to choose English or Hybrid over Leb Lango. Specifically, for each standard deviation increase in Week 1 score, the odds of selecting English instead of Leb Lango increased by 49\% ($b = 0.40$, $OR = 1.49$, $p < .0001$), and the odds of selecting Hybrid instead of Leb Lango increased by 50\% ($b =.40$, $OR = 1.50$, $p < .0001$). These results indicate that learners with stronger early performance tended to select English or Hybrid instruction, whereas learners with lower baseline scores were more likely to choose Leb Lango.

\subsection{RQ3: How do learners who selected Leb Lango, English, or Hybrid instruction engage with the lesson content?}

\begin{figure}[h] 
\centering \includegraphics[width=\linewidth]{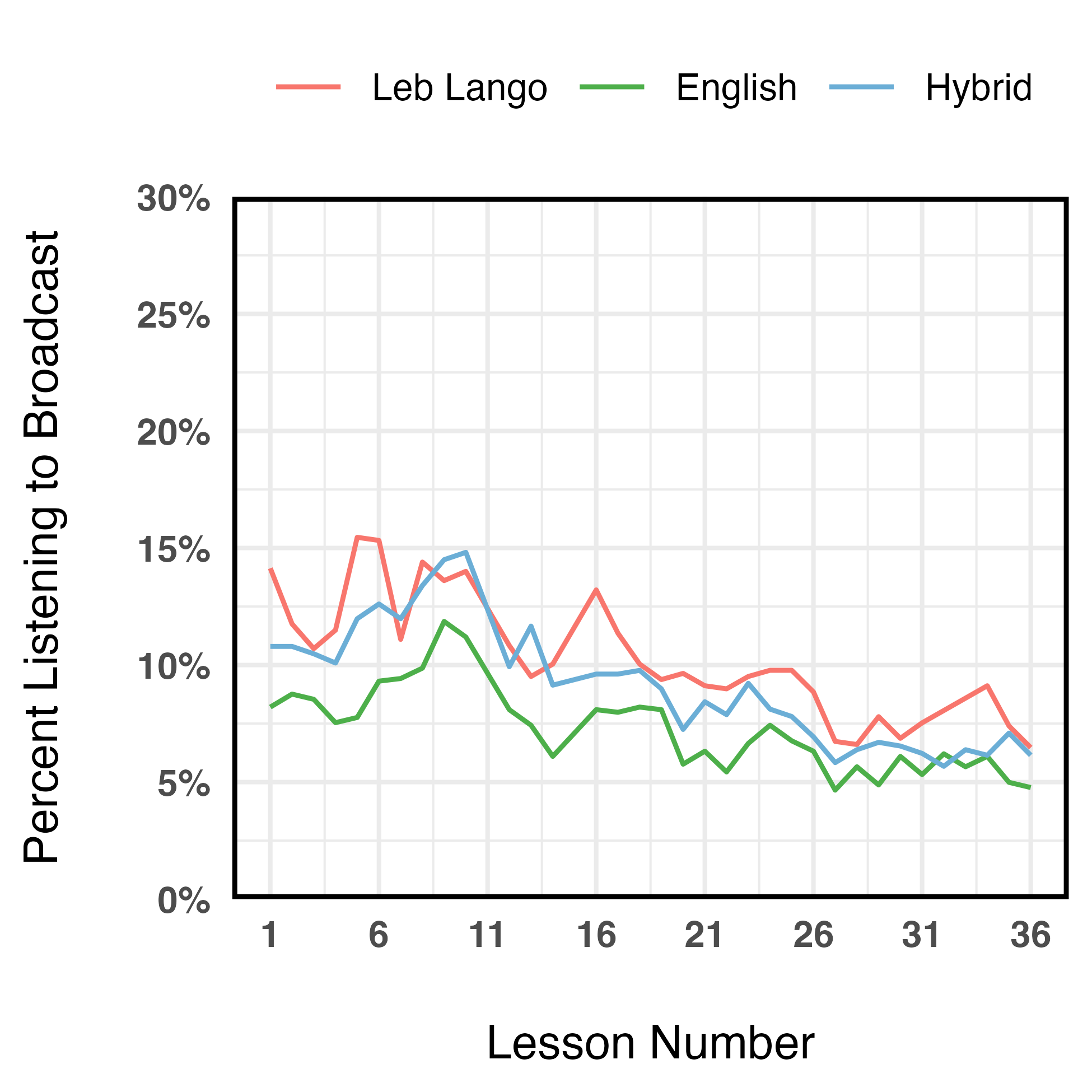} 
\caption{Percent of learners listening to live radio broadcasts (English and Leb Lango broadcasts) by language option of the course (Leb Lango, Hybrid and English). Each broadcast was associated with a distinct prize code, and learners’ submitted codes were used to classify which broadcast they listened to. The x-axis represents lesson number, while the y-axis shows the percentage of learners engaged with each lesson.}
\Description{This figure shows the percentage of learners listening to radio broadcasts for each version of the course, that is English and Leb Lango, for each language choice (Leb Lango, Hybrid and English). The x-axis represents lesson number, and the y-axis shows the percentage of learners engaged with each lesson.}
\label{fig:engagement1} 
\end{figure}

To measure engagement, we examined how learners in the filtered sample of 1,326 participants across the three course language options (Leb Lango, English, and Hybrid) engaged with live radio broadcasts, using the accuracy of the prize code as the engagement measure \cite{asher_validating_2025}. The broadcasts themselves were delivered in either English or Leb Lango. The learners in the Hybrid group had the flexibility to listen to either broadcast, depending on the language they selected at the beginning of each step. Because instructors for the English and Leb Lango broadcasts used distinct prize codes, we also classified which broadcast each learner listened to based on the code they entered. Figure ~\ref{fig:engagement1} shows how frequently the learners in each version of the course submitted a correct prize code for each lesson - our measure of whether the learners were listening to each broadcast. 

Overall, engagement was relatively low, with fewer than 14\% of learners submitting a correct prize code in any given week. However, engagement patterns varied across language groups. From the beginning of the course, learners enrolled in the English version of the course were significantly less likely to tune in for live broadcasts than those enrolled in the Leb Lango version ($OR = .47$, $z = -4.40$, $p < .001$), indicating higher participation in live radio broadcast among Leb Lango learners than among English learners. In contrast, learners engaged with the radio broadcasts at similar rates in the Leb Lango and Hybrid versions of the course ($OR = .90$, $z = -0.70$, $p = .484$). Overall, engagement with broadcasts declined over time ($OR = 0.88$, $z = -16.24$, $p < .001$), pattern typical of remote learning settings. The rate of decline did not differ significantly across the three language groups ($ps > .378$), suggesting that the temporal drop-off was consistent across language groups.

\begin{figure}[h] 
\centering \includegraphics[width=\linewidth]{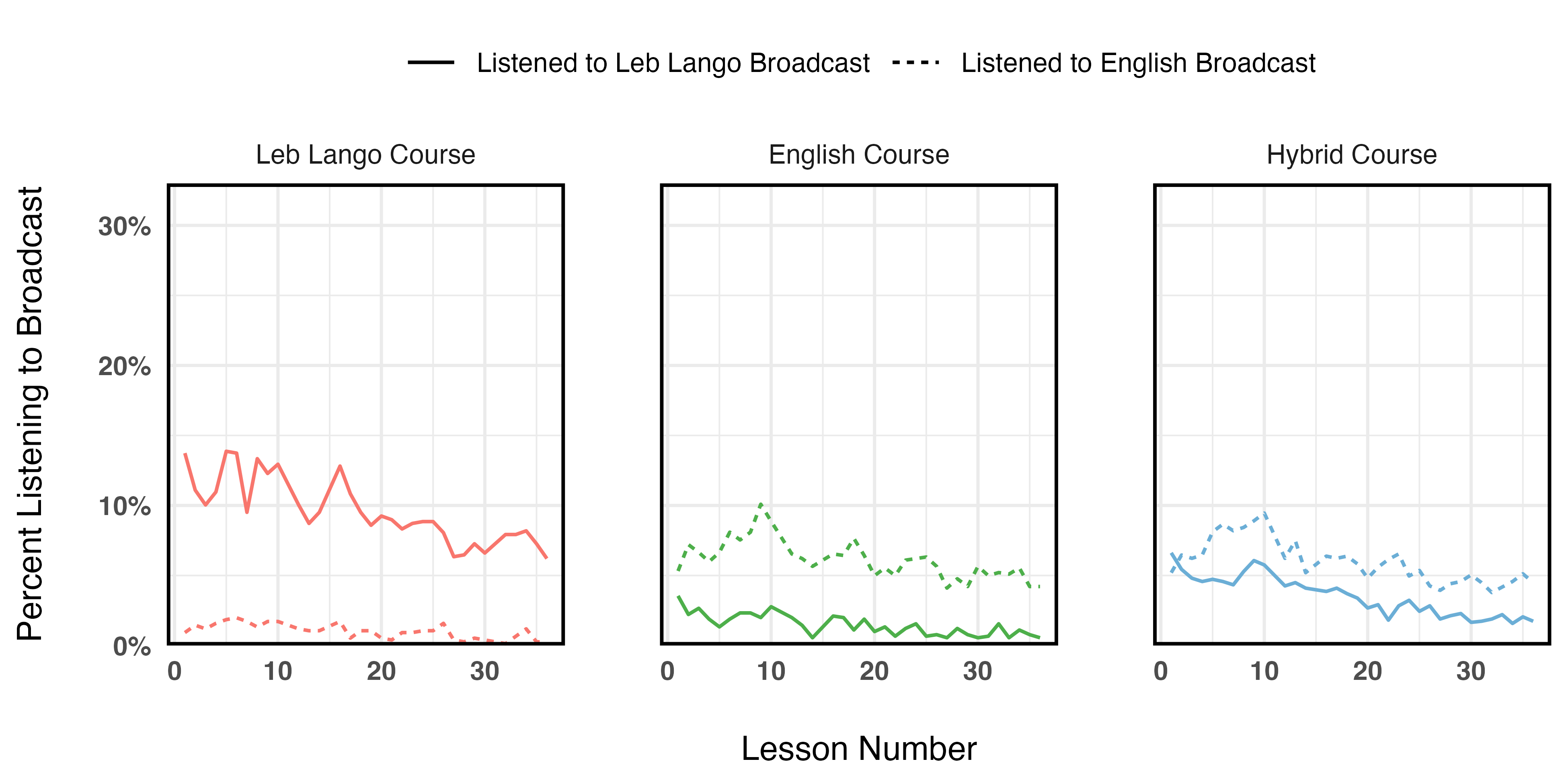} 
\caption{Percent of learners listening to the Leb Lango radio broadcast versus the English radio broadcast for each language choice.}
\Description{This figure shows the percent of learners listening to the Leb Lango versus English radio broadcasts for each language choice (Leb Lango, Hybrid and English). The x-axis represents lesson numbers, and the y-axis shows the percentage of learners engaged with each lesson.}
\label{fig:engagement2} 
\end{figure}

Next, to understand which specific version of the radio broadcast learners listened to, we examined the rate at which learners in each language option submitted the prize codes read aloud in the English broadcasts (aired at 12 PM) versus the codes read aloud during the Leb Lango broadcasts (aired at 2 PM). The results of this analysis are presented in Figure ~\ref{fig:engagement2}.

In both the Leb Lango and English versions of the course, learners favored the radio broadcast corresponding to their preferred language option. This was particularly the case in the Leb Lango version of the course, where learners were 9.5 times more likely to engage with the Leb Lango broadcast, remaining largely in the intended language option. In the English version of the course, learners were slightly more evenly distributed between broadcasts, favoring the English broadcast at a ratio of 4.2 to 1, while a subset also accessed the Leb Lango broadcast. This cross-language engagement is evident among learners in both the English and Leb Lango versions of the course, but it is more pronounced among English-enrolled learners, suggesting that used the local language broadcast as an additional form of linguistic support. In the Hybrid version of the course, learners used the English broadcast 1.7 times more frequently than the Leb Lango broadcast. 


\subsection{RQ4: How do learners who selected Leb Lango, English, or Hybrid instruction progress through the course?}
\FloatBarrier
 
Next, we investigated learning trajectories over time. Examining trajectories is important because it allows us to understand not just how well learners perform overall, but how they progress over time, particularly relevant in this context, where initial analyses showed that learners in the Leb Lango group began the course with lower baseline scores. Modeling these trajectories helps determine whether these early gaps persist, narrow, or widen over time. 

To examine learning trajectories across language groups, we fitted a linear mixed-effects model (lme4 in R) on the filtered sample of 1,326 participants with average score as the dependent variable. We fit linear mixed-effects models separately for practice assessments and weekly tests because they capture different aspects of learning using the following:
\[
\text{average score} \sim \text{week} + (\text{week} \,|\, \text{language}) + (1 \,|\, \text{user id})
\]
Here, we modeled the average scores of the learners using the week in the course as a fixed effect to capture general performance trends (\text{week}). To allow learning trajectories to vary by course language, we included a random effect of week nested within course language ($(\text{week} \,|\, \text{language})$), letting both intercepts and slopes differ across language groups. In this context, the intercept represents the average starting performance of learners at the beginning of the course, while the slope represents the rate of change in performance over time (i.e., the learning trajectory). This approach accounts for the fact that different language groups may start at different levels of performance and learn at different rates. To account for repeated measurements from the same learners, we also included a random intercept for each learner ($(1 \,|\, \text{user id})$), which captures unobserved differences in baseline performance and addresses the non-independence of observations within individuals.

Table ~\ref{tab:lme_results} summarizes the estimated intercepts and slopes. Consistent with the results of RQ2, for practice tasks, we find that higher baseline performance, as estimated by this model, is associated with choosing English or Hybrid (69.5\% and 73.2\% percent correct) compared to choosing Leb Lango (49.5 \% percent correct).

Moreover, we see that the week-to-week increase (slope) for practice tasks is greater for learners who chose Leb Lango ($0.040$) than for learners who chose English ($0.001$) or Hybrid ($-0.007$) instruction. The weekly tests showed a similar pattern, with faster improvement among Leb Lango learners than English or Hybrid learners, despite their lower initial performance.

Overall, across both practice tasks and weekly tests, learners who began with lower initial performance in Leb Lango showed faster learning rates over time, while those in English and Hybrid started with higher performance but improved more slowly. 

\begin{table}[h]
  \centering
  \caption{Estimated intercepts and slopes from linear mixed-effects models for practice assessments and weekly tests by course language option. Intercepts indicate initial performance levels, and slopes indicate weekly learning rates.}
  \Description{This table shows results from a linear mixed-effects model, displaying intercepts and slopes for different language options (Leb Lango, English and Hybrid). Intercepts represent starting points and while slopes represent weekly growth rates.}
  \label{tab:lme_results}
  \begin{tabular}{lccc}
    \toprule
    \textbf{Assessment} & \textbf{Language choice} & \textbf{Intercept} & \textbf{Slope} \\
    \midrule
    \multirow{3}{*}{Practice}
      & Leb Lango & 49.5\% & $0.040$ \\
      & English   & 69.4\% & $0.001$ \\
      & Hybrid    & 73.2\% & $-0.007$ \\
    \midrule
     \multirow{3}{*}{Weekly tests} 
      & Leb Lango & 46.8\% & $0.111$ \\
      & English   & 64.1\% & $0.045$ \\
      & Hybrid    & 66.4\% & $0.035$ \\
    \bottomrule
  \end{tabular}
\end{table}

\subsection{RQ5: How do learners who selected Leb Lango, English, or Hybrid instruction perform in the course?}

\begin{figure}[htbp!] 
\centering \includegraphics[width=\linewidth]{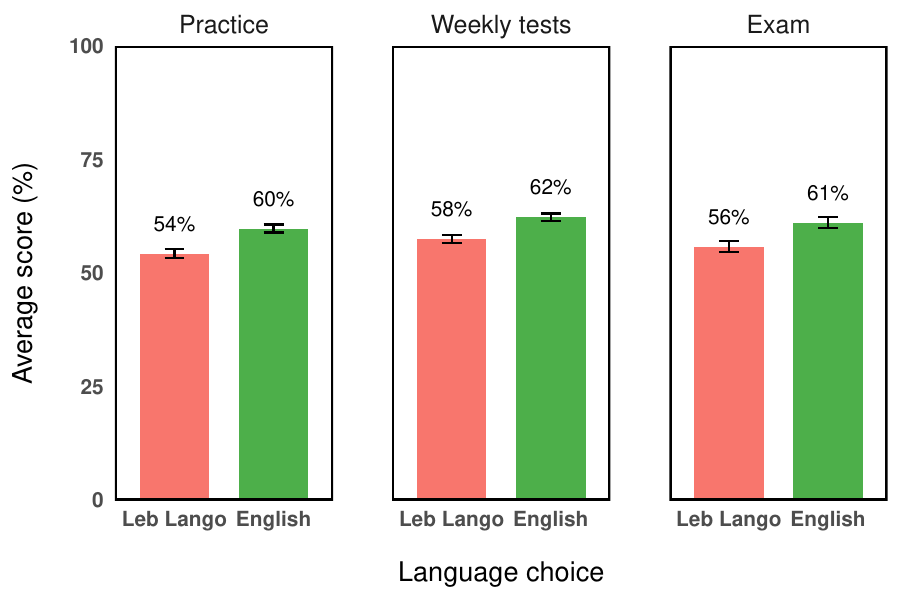} 
\caption{Comparison of learners’ average scores in Leb Lango and English across practice tasks, weekly tests, and final examination.} 
\Description{This figure compares learners’ average scores in Leb Lango and English across different assessment types, including practice tasks, weekly tests, and final exams on a matched sample.}
\label{fig:rq5_1} 
\end{figure}

Finally, we investigated the relationship between language preference and learner performance. To address missing data and ensure fair comparisons across language groups, we restricted the sample to learners who attempted at least eight course steps and completed the introductory step test in the first week of the course. This resulted in a final sample of $1,262$ learners (273 in Leb Lango, 421 in English, and 568 in the Hybrid condition). To account for baseline differences, we applied propensity score matching (PSM) based on Week 1 tests performance, conducting two pairwise comparisons with Leb Lango as the treated group and English and Hybrid as the control groups: Leb Lango vs. English and Leb Lango vs. Hybrid. We evaluated multiple matching algorithms and selected nearest-neighbor matching with mahalanobis because it achieved the best improved covariate balance, lower mean distances between matched pairs and higher sample retention. For both comparisons, the matched samples included 273 learners in each group; retaining 100\% from Leb Lango, 65\% from English and 48\% from Hybrid. Although some control units were lost during matching, the overall balance improved and met commonly accepted standards for adequate covariate balance \cite{zhang_balance_2019}: for the Leb Lango vs. Hybrid comparison, the standardized mean difference decreased from $0.371$ to $0.003$; for the Leb Lango vs. English comparison, the standardized mean difference decreased from $0.368$ to $0.067$. After matching, we fitted linear regression models for each assessment type for the separate comparisons: practice, weekly tests, and final examination, with demographic variables (educational level and neighborhood type) and Week 1 performance included as covariates. We present the learners' average scores for English vs. Leb Lango and Hybrid vs. Leb Lango in Figure ~\ref{fig:rq5_1} and Figure ~\ref{fig:rq5_2} respectively, across practice assessments, weekly tests, and the final examination. 
\subsubsection{Leb Lango and English:}
\FloatBarrier
After matching learners on Week 1 performance and controlling for both demographics and Week 1 performance, we found that for the practice assessments, language choice was a significant predictor of practice assessment performance. Leb Lango learners scored significantly lower than English learners on the practice assessments (54\% vs. 60\%; $b = -0.240$, $p < .01$). However, the association of Leb Lango choice with practice scores is smaller in magnitude than the association of starting with a higher baseline or having a higher education level. Week 1 baseline scores were the strongest predictor of practice performance ($b = 0.467$, $p < .001$), indicating that learners with higher starting scores tended to perform better in practice assessments than learners with lower starting scores. The school level was positively associated with the average practice scores: learners with the S1-S4 schooling (Lower secondary) scored higher than those with P1–P4 ($b  = 0.339$, $p = .015$), and S5+ was not significantly associated practice performance ($b = 0.353$, $p = .061$). These results suggest that learners in lower secondary performed better on practice tasks, while learners in primary schooling or higher secondary schooling did not differ significantly. Neighborhood type was not a statistically significant predictor of practice scores ($b = -0.139$,\ $p = .082$).

On the weekly tests, Leb Lango learners scored 58\% on average compared to 62\% for English learners. However, language choice was not a significant predictor of weekly test performance ($b = -0.146$, $p = .064$). As in the practice model, week 1 score  showed a strong positive association with the weekly test scores ($b = 0.481$, $p < .001$). Schooling level exhibited an even stronger association with performance. Schooling was the strongest predictor of weekly test scores: learners with S1–S4 (Lower secondary) ($b = 0.501$, $p < .001$) and S5+ (Upper secondary) ($b = 0.647$, $p < .001$) were more likely to outperform those with P1–P4 (Lower primary). Furthermore, learners living in urban areas and rural areas performed similarly on weekly tests, with those living in urban areas slightly doing worse ($b= -0.148$,\ $p = .055$).

\FloatBarrier
Similarly, on the final exam, Leb Lango learners scored 56\% versus 61\% for English learners, with the difference also not significant ($b = -0.065$, $p = .492$),  suggesting that by the final exam, differences due to assessment had largely diminished. The association between schooling and final exam performance was substantial and continued to show a clear gradient: learners with upper primary schooling performed better on the final exam ($b = 0.322$, $p < .05$), and those with lower secondary schooling ($b = 0.689$, $p < .001$) and with upper secondary schooling ($b = 1.042$, $p < .001$) scored significantly higher than lower primary schooling learners. Week 1 score was a strong predictor of final exam scores ($b = 0.375$, $p < .001$). The urban/rural association was again negative but not a significant predictor ($b = -0.059$,\ $p = .519$).

Across all assessments, we see that learners who chose Leb Lango were associated with lower performance early on but later performed similarly to English learners. Furthermore, learners’ Week 1 scores and schooling level were stronger predictors of performance than language. Week 1 score was the most consistent predictor of performance across all assessments, with higher starting scores associated with better learning outcomes. Schooling level was positively associated with performance, with learners at higher levels of schooling generally achieving better scores, particularly on weekly tests and the final examination.

\FloatBarrier
\subsubsection{Leb Lango and Hybrid:}
\begin{figure}[H] 
\centering \includegraphics[width=\linewidth]{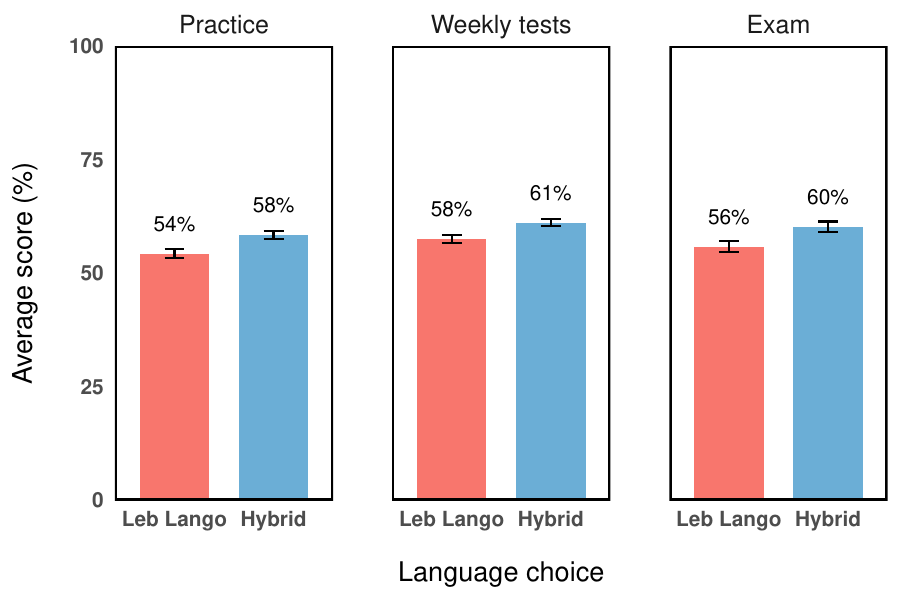} 
\caption{Comparison of learners’ average scores in Leb Lango and Hybrid across practice tasks, weekly tests, and final examination.}
\Description{This figure compares learners’ average scores in Leb Lango and Hybrid across different assessment types, including practice tasks, weekly tests, and final exams on a matched sample.}
\label{fig:rq5_2} 
\end{figure}

Figure ~\ref{fig:rq5_2} and the model results reveal similar patterns when comparing Leb Lango and Hybrid learners. On the practice assessments, Leb Lango learners scored 54\% on average, performing significantly lower than Hybrid (58\%) learners ($b = -0.205$, $p < .01$). Week 1 score was positively associated with practice scores ($b = 0.023$, $p < .001$). The largest association with practice performance was observed for lower secondary level (S1–S4): learners with S1–S4 scored significantly higher than P1–P4 learners ($b = 0.416$, $p < .01$). Learners with P5–P7 ($b = 0.206$, $p = .117$) and S5+ (Upper secondary) ($b = 0.303$, $p = .137$) showed insignificant associations with practice performance. Neighborhood type did not significantly predict practice scores ($b = -0.129$, $p = .113$).

On the weekly tests, Leb Lango learners scored significantly lower, 58\%, compared to 61\% for Hybrid learners ($b = -0.181$, $p < .01$). Week 1 score was positively associated with weekly test scores ($b = 0.025$, $p < .001$). Again, lower secondary schooling had the largest association with weekly test performance. Learners with S1–S4 ($b = 0.394$, $p = .003$) scored higher on weekly tests than P1-P4 learners, while P5–P7 ($b = 0.128$, $p = .316$) and S5+ ($b = 0.310$, $p = .118$) were not significantly different. Neighborhood type showed an insignificant negative association ($b = -0.136$, $p = .086$).

On the final examination, Leb Lango learners scored 56\% versus 60\% for Hybrid learners, though this difference was not statistically significant ($b = -0.107$, $p = .245$). Week 1 score significantly predicted final exam performance ($b = 0.019$, $p < .001$). Schooling showed a positive significant association with a clear positive gradient: learners with P5–P7 ($b = 0.373$, $p < .05$), S1–S4 ($b = 0.646$, $p < .001$), and S5+ ($b = 0.751$, $p < .01$) scored higher than P1–P4 learners. Neighborhood type was not a significant predictor ($b = -0.055$, $p = .558$).

Overall, these results show a trend that suggests that learners who chose Leb Lango scored lower than Hybrid learners early on, but this gap narrowed by the final examination. Schooling level and Week 1 scores were consistent predictors of learners' performance. Notably, schooling level—especially lower secondary education—showed the strongest and most consistent association with performance across all assessment types.

\section{Discussion}

In a typical multilingual classroom, it is not feasible for teachers to deliver lessons in more than one language simultaneously, and learners do not have the option to choose their language of instruction. Our study provides important insights into how the affordances created by a multilingual digital learning environment relate to patterns of access, engagement, and learning in mobile learning contexts. The availability of a local language option appears to open participation, particularly among learners with lower prior knowledge, lower levels of formal schooling, and those from rural areas. Local language choice was also associated with higher engagement and faster learning patterns for these learners, underscoring linguistic accessibility as a critical design affordance in low-tech mobile learning environments. Our findings build on HCI and EdTech research showing that mobile learning can expand educational access for marginalized learners, and extend this research by demonstrating how language choice itself functions as a usability and accessibility mechanism. Language is one of the most foundational components of usability, and our work emphasizes that meaningful learning interactions can emerge when learners navigate course content in a language they understand. We discuss these findings in greater detail below.

\subsection{Designing to Expand Access for Disadvantaged Learners through Local Language Choice}
We set out to understand the extent to which providing low-tech mobile learning in a local language can broaden student participation in a non-formal remote STEM education. Our results show that learners who chose the local language option were more likely to have lower levels of education, live in rural areas, or have lower baseline scores. This pattern highlights that local language provision is linked to greater participation among learners who are typically disadvantaged. More broadly, the results suggest that delivering instruction in learners’ mother tongue or a familiar language could, over time, enable additional learners to engage in learning opportunities that might otherwise remain inaccessible.
This relationship is particularly salient in non-formal remote settings, where learners may not have direct access to teachers or structured scaffolding, and must navigate the material independently. In such contexts, language options provided by technology serve as a key design lever to reduce the additional barrier to entry and give learners the agency to engage with the course through the language they find most accessible. Without the inclusion of local language options, reliance on monolingual instruction, which is very common in mobile learning, may risk excluding learners who do not speak the language of instruction.

Our findings align with patterns observed in regional MOOCs such as Edraak and XuetangX, where lower-educated learners were found to be more represented in courses taught in the local language \cite{ruiperez-valiente_large_2022}. However, these studies primarily focused on tertiary education and categorized K–12 learners into broad categories. Building on this work, our study provides a more detailed perspective by disaggregating K–12 learners and specifically examining the preferences of those with very limited prior schooling (P1–P4) in a non-formal learning context.

Patterns of language choice in our data may reflect differences in learners’ language exposure—shaped by factors such as geography and education policy—and help contextualize which learners might benefit most from local-language inclusion in educational technology. The association between the highest education level and language option in this context likely reflects differences in prior exposure to English, consistent with Language-in-Education Policy (LiEP), which mandates mother tongue instruction in lower primary (P1–P3) followed by transition to English as the language of instruction in P4 in government schools \cite{altinyelken_dilemmas_2014}. Therefore, learners whose highest education level is lower primary (P1-P4) are still primarily educated in their mother tongue or at the transition phase to English, and are less comfortable navigating English-based mobile learning environments. In contrast, those who have attained higher schooling levels have accumulated greater exposure to English, which may have facilitated their ability to engage with English or Hybrid systems that include English. Notably, these education policy effects are unevenly implemented. Prior research shows that mother-tongue instruction is applied more consistently in rural schools and has a stronger impact on educational outcomes for children in rural areas than in urban areas \cite{seid_does_2016}. Urban learners, meanwhile, tend to have greater exposure to English through schooling, media, and everyday interactions \cite{seid_does_2016, ssentanda_challenges_2014, mncwango_language-related_2021}. Accordingly, rural learners in our study were more likely to choose Leb Lango, whereas urban learners often opted for English or Hybrid instruction. Given these rural–urban differences in LiEP implementation, language choice in EdTech holds potential to operationalize policy goals.

Learners’ prior knowledge also varied across the language groups, with learners who had lower prior knowledge more likely to select Leb Lango. Prior research shows that prior knowledge is a significantly positive predictor of student achievement \cite{thompson_prior_2003, hammadou_interrelationships_1991}, and STEM content is often complex and largely inaccessible to disadvantaged learners \cite{makgato_factors_2007, visser_home_2015}. In this context, mobile learning in Leb Lango reached learners with limited prior STEM knowledge and provided a more accessible entry point to engage with challenging STEM material that might have been more difficult to navigate through English-only instruction.   

Furthermore, it is worth noting that many learners in our study chose English or Hybrid instruction, with the majority (43\%) selecting the Hybrid option that allowed them to switch between Leb Lango and English. The high uptake of the Hybrid option reflects learners’ desire to leverage both Leb Lango and English when engaging with the course content. This further highlights how linguistic choice in EdTech can serve as a mechanism for improving usability and accessibility, enabling participation among learners who may have varied language proficiencies, diverse needs, and motivations. While our study did not investigate why learners made these language choices, understanding their preferences is essential for EdTech design. Future research could explore the reasons behind these preferences, providing insights that help create learning experiences accommodating diverse learner priorities: some may prioritize comprehension, others may value the prestige or utility of a particular language \cite{kwon_navigating_2025}, and others may seek the flexibility to switch between languages.

\subsection{Local Language Radio Lessons as an Accessibility Affordance for Supporting Engagement} 

Engagement with radio broadcast lessons was generally low across all language options, reflecting the challenges of low-tech remote learning. Several contextual factors likely contributed to this pattern. Although mobile phones are widely accessible, some learners, for example, do not own personal devices and must share phones with family members, which can limit their ability to participate in activities such as entering prize codes and, in turn, reduce overall engagement. It is important to note that because radio broadcasts are openly accessible and do not restrict who can listen, learners were able to tune in to broadcasts of any language regardless of their initial registration choice. This provided nuanced insight into learner behavior, revealing both those who consistently engaged with their selected language and those who explored alternative language options. 

Our findings indicate that engagement with radio broadcasts declined over time at a similar rate across all language groups, showing that temporal drop-off was consistent regardless of language choice. While the course  expanded access to a more diverse learner population, sustaining engagement over time in a low-tech environment may require additional design considerations beyond language.
Despite similar patterns of decline, important differences emerged across language groups. We found that learners enrolled in the English course were significantly less likely to tune in for live broadcasts than those enrolled in the Leb Lango version. This pattern suggests that Leb Lango instruction was associated with greater overall engagement compared to English, even as temporal decline occurred similarly across groups. Higher broadcast participation among Leb Lango learners may reflect the oral nature of local languages, which are primarily spoken and learned through listening and community interaction in many African contexts \cite{abdi_oral_2007}. Radio broadcasts in Leb Lango may therefore have aligned with these oral practices, providing an accessible and culturally familiar mode of learning and engagement. Consistent with these findings, we found that learners across the three language choices favored the radio broadcast corresponding to their preferred language; however, learners who registered for the Leb Lango course showed greater consistent alignment with the intended broadcast than those in English, likely because course concepts were accessible in a more familiar language. 

At the same time, we observed cross-language engagement among Leb Lango and English learners, with some participants accessing broadcasts in a language they had not initially registered for. Cross-language engagement pattern was more pronounced in the English group, suggesting that local language broadcasts provided alternative or additional access and comprehension support, even for those who initially chose English. This finding aligns with prior studies showing that learners often revert to their home languages in English-instructed classroom discussions while remaining highly engaged and on-task \cite{msimanga_talking_2014}. In our context, learners may have drawn on Leb Lango as a resource when encountering comprehension difficulties, thereby lowering barriers to participation. These observations emphasize the role of multilingual EdTech tools in supporting participation and engagement. Our findings contribute new insights into how learners engage with multiple languages through audio or radio modalities in mobile technologies. Overall, incorporating Leb Lango into the EdTech remote course supported engagement among learners who selected the language, while the availability of multiple language options enabled learners who chose Leb Lango and English to draw flexibly on linguistic resources based on their learning needs.

\subsection{Language Affordances Supporting Learning Gains over Time and Reducing Performance Gaps}
Our results suggest that providing instruction through mobile learning in a local language chosen by learners as the one most familiar to them is associated with greater learning gains, particularly for learners who started at lower levels of knowledge. Learners who chose English or Hybrid instruction began the course with higher initial performance, which may reflect earlier exposure to English-medium schooling or broader educational opportunities. In contrast, learners who selected Leb Lango started with lower baseline scores. Despite these initial disparities, the trajectory of learning shows that, across both practice tasks and weekly tests, learners in the Leb Lango group demonstrated faster improvement than learners in the English or Hybrid groups. This pattern suggests that instruction delivered in a local language can help learners with weaker starting points learn more and gain momentum as they progress through the course.

At the same time, analyses of the relationship between language choice and performance reveal a more nuanced pattern. Our results showed that learners who chose Leb Lango were associated with lower practice scores and weekly test scores than English or Hybrid instruction. Nevertheless, by the final examination, performance differences between Leb Lango and English learners, and Leb Lango and Hybrid learners, narrowed and were no longer significant. While our findings differ from research showing that learners taught in their mother tongue often demonstrate stronger learning outcomes \cite{piper_implementing_2016, trujillo_use_2020}, they offer insight into how language-related disadvantages may interact with other contextual factors: Leb Lango learners generally had less prior schooling, lower initial knowledge, and were more likely to come from rural areas. Thus, their lower initial scores likely stemmed from limited prior knowledge or other unmeasured factors, such as difficulty comprehending complex academic content, limitations in the translation of scientific terminology, or the possibility that Leb Lango was not the learners’ most typical home language, even if it was the most familiar option available. 
Importantly, while offering the course in Leb Lango was associated with faster learning rates, it did not fully compensate for gaps in foundational knowledge that existed prior to the course or other unmeasured factors that shaped early performance. In fact, a key pattern emerges when examining performance beyond language choice: while choosing Leb Lango was associated with lower performances in practice and weekly tests relative to learners using English or Hybrid, this association was not the strongest factor explaining performance differences. Instead, baseline knowledge and schooling level were the strongest factors. Learners with higher prior knowledge showed higher performance, and learners with more schooling performed better than those with less schooling. While EdTech can provide critical support for access and engagement through language choice, additional interventions addressing disparities in foundational knowledge and prior schooling are necessary to help learners achieve stronger learning outcomes and fully benefit from language-inclusive instruction.

\section{Implications}

Our study highlights several practical considerations for designing educational technologies in multilingual contexts. First, we demonstrate that enabling learners to engage in their preferred language is both feasible and effective, even in low-tech environments such as radio- and phone-based learning systems. Multilingual digital learning technologies are uniquely positioned to scale instruction across multiple languages in ways that traditional classrooms cannot, reaching learners disadvantaged by limited formal schooling or exposure to dominant languages.

However, our findings suggest that language support alone is insufficient to address all learning barriers. Other factors such as differences in prior knowledge and schooling remain important. In practice, this suggests that multilingual EdTech should be paired with targeted additional support. In low-tech contexts, this could take the form of contextualized examples, hints delivered via SMS, or step-by-step audio guidance. For example, learners with lower prior schooling may need more frequent or detailed  step-by-step audio guidance at the beginning of the course. By combining language accommodations with targeted supports, EdTech designers and instructors can better ensure equitable learning outcomes.

While our study explored three language options, it also highlights opportunities for EdTech designers to scale similar approaches in other contexts and support a larger set of languages. Scaling to multiple languages may increase system complexity, so designing such systems requires careful attention to navigation and usability. This will require interfaces with clear and recognizable labels, straightforward menu placement, easy access to language options, and audio cues to guide users intuitively. In low-tech contexts, EdTech designers must carefully balance usability with technological constraints to ensure learners can access content reliably in their preferred language.

The cross-language engagement observed in our radio broadcasts underscores the value of flexible, multimodal multilingual interaction models. Future EdTech systems, especially those incorporating local languages with rich oral traditions and limited written documentation, may benefit from leveraging audio-based explanations or hybrid audio-text approaches to support comprehension. In addition, such systems can generate valuable data to inform the development of machine learning models for low-resource languages, ultimately improving the design and effectiveness of multilingual EdTech.

Dynamic and adaptive language support represents a promising frontier for future EdTech systems. In our study, we observed cross-language engagement among learners. This raises critical questions regarding how systems should effectively detect, respond to, and support learners’ shifting language preferences in real-time. Realizing this vision will depend on advancements in language technologies and data-driven approaches, as well as collaboration with language experts, particularly for the low-resource languages that lack sufficient data. With these technologies in place, designers can explore mechanisms that allow learners to switch languages mid-task, access on-demand translations, or receive mixed-language explanations.

Finally, to continue advancing multilingual EdTech, more efforts should be made to conduct empirical studies evaluating these systems and to incorporate feedback from learners about their experiences with multilingual EdTech, so that the resulting systems are learner-centered and grounded in learners’ experiences and realities.

\section{Limitations}

Our study has several limitations. Firstly, we employed a quasi-experimental design; consequently, the absence of randomization introduces potential self-selection bias. Learners who chose a particular language may have differed systematically from those who did not. We mitigated this to some extent using propensity score matching. While the matching process improved the balance between groups, it also reduced the sample size and may have affected statistical power. 

Secondly, due to platform constraints such as character limits and the large-scale nature of the study, we could not collect richer demographic information or baseline assessment data. Factors such as prior exposure to Leb Lango, home language use, and other unmeasured variables may have influenced both language choice and performance. In place of formal baseline assessments, we used Week 1 performance as a proxy for prior knowledge, which may not fully capture initial differences in preparedness. Moreover, in a low-resource context such as ours, conducting a formal baseline study might have discouraged learner participation and ultimately reduced access, undermining the very goals of the EdTech intervention. Furthermore, because the system relied on USSD, we were constrained to multiple-choice items, which narrowed the range of possible question types and limited the insights we could gather about how learners use language and interact with the technology.

Thirdly, we did not systematically assess the quality or accessibility of the translations, nor did we examine whether the English and Leb Lango versions were equivalent in difficulty. Translation choices and potential differences in item difficulty may have influenced how learners navigated the course and engaged with the content. Consequently, observed differences in performance between language groups should be interpreted with this potential source of variation in mind.

Finally, we could not account for variability in instructional delivery, including differences in radio broadcast quality or instructors' proficiency in each language, both of which could have influenced learning outcomes.

While these limitations should be considered when interpreting our findings, our study provides valuable insights into how multilingual EdTech with learner-driven language choice can expand access and participation.

\section{Conclusion and Future Work}
Our study provided the first large-scale evidence on multilingual mobile learning through language choice. We demonstrated how mobile technologies that incorporate local languages can expand access and support learning in resource-constrained, non-formal settings. We conducted a quasi-experimental study in which learners self-selected their preferred language for a STEM course: English, Leb Lango (a local language), or a combination of both Leb Lango and English. Our findings show that offering a local language option was associated with broader participation, particularly among learners with limited prior STEM knowledge, lower levels of formal education, and those from rural areas—groups that may be underserved by English-only instruction. Learners who selected Leb Lango were also more actively engaged. Although they began with lower performance, they improved faster than English and Hybrid learners, ultimately closing initial performance gaps by the final examination. 

We conceptualize language choice in EdTech as a personalization mechanism that expands educational opportunities and fosters more inclusive learning pathways for disadvantaged learners. The benefits of mother-tongue instruction documented in classroom-based research may extend to mobile and technology-mediated learning environments and, importantly, have the potential to reach larger populations through digital platforms. 

Building on these insights, future research will aim to provide causal evidence on the effects of language of instruction in mobile learning through a randomized controlled trial. Further work is also needed to examine the quality and consistency of translations for scientific and technical content, as well as the relative difficulty of instruction across languages, to ensure meaningful comparability. Finally, future work should complement quantitative data with interviews to understand learners’ motivations for language preference, as well as learners' and instructors' experiences and perceptions of multilingual mobile learning, in order to surface nuanced insights that inform more effective and context-sensitive multilingual EdTech design.

\begin{acks}
This research was supported by the Jacobs Foundation (Award No. 2023151900) and the Mastercard Foundation. We thank our collaborator, Yiya Solutions, Inc., for providing access to the data and providing extensive support throughout this research. Finally, we sincerely thank the students and instructors who participated in the course studied.
\end{acks}
\bibliographystyle{ACM-Reference-Format}
\bibliography{references}
\end{document}